\documentclass[twocolumn,showpacs]{revtex4}
\usepackage{epsfig}

\begin{document}

\author{E. Barkai}
\affiliation{Department of Physics,
Bar Ilan University, Ramat-Gan 52900 Israel, e-mail: barkaie@mail.biu.ac.il}

\author{I.M. Sokolov}
\affiliation{Institut f\"ur Physik,
Humboldt-Universit\"at zu Berlin, Newtonstra\ss e 14, 12489 Berlin, Germany;
e-mail: igor.sokolov@physik.hu-berlin.de}

\title{Multi-point Distribution Function for the Continuous Time Random Walk}

\pacs{05.40.Fb, 02.50.-r}

\begin{abstract}

We derive 
an explicit expression for the Fourier-Laplace transform
of the two-point distribution function $p(x_1,t_1;x_2,t_2)$
of a continuous time random walk (CTRW),
 thus generalizing the result of Montroll 
and Weiss for the single point distribution function $p(x_1,t_1)$.
The multi-point distribution function has a structure of a 
convolution of the Montroll-Weiss CTRW and the aging CTRW 
single point distribution functions. The correlation function $\langle x(t_1) x(t_2) \rangle$
for the biased CTRW process is found. 
The random walk foundation of the multi-time-space fractional 
diffusion equation 
[Baule and Friedrich [{\em Europhysics
Letters} {\bf 77} 10002 (2007)] 
is investigated using the unbiased CTRW in the continuum limit.
\end{abstract}

\maketitle

\section{Introduction}
The continuous time random walk (CTRW) introduced in \cite{MW} more
than forty years ago is a successful model for normal and anomalous
diffusion in a variety of physical systems
\cite{Kehr,BouchaudREV,Metzler,Zaslavsky,Flom}. A new splash of
interest in this old model was caused by the fact that it provides a
firm statistical foundation of the fractional Fokker-Planck equation
\cite{FFPEPRL,BarkaiPRE}, and is a simple model for the investigation
of such intriguing phenomena of non-equilibrium statistical physics as
weak ergodicity breaking \cite{Bel} and statistical aging
\cite{ACTRW,ACTRWPRL,SokolovPRL,ecm}.  The importance of CTRW as a minimal
model of non-Markovian behavior is connected with its semi-Markovian
(renewal) nature, which allows in many cases for an exact
probabilistic description of the process of interest.  Unlike
Markovian diffusion processes, which are fully characterized by their
transition probabilities, non-Markovian CTRW requires for the full
hierarchy of multi-point distribution functions for its complete
characterization
\cite{Grigolini,Barsegov,Mukamel,BauleEPL,BaulePRE}. Multi-point
distribution functions and correlation functions are necessary tools
to distinguish between CTRW stochastic dynamics from other non
Markovian processes, e.g. fractional Brownian motion, as appearing in
linear fracton models or in generalized Rouse models, see
e.g. \cite{GraKla}. The simplest experimental way of discriminating
these types of anomalous processes is based on different behavior of
their correlation functions, as investigated e.g. in recent
experiments on anomalous diffusion in single molecules
\cite{Yang,Min,Margolin}.  Hence obviously it is a worthy goal to
consider higher-order joint probability density functions (PDFs) and
correlation functions of the CTRW.

In the present article we concentrate on decoupled CTRWs in the
sub-diffusive and spatially homogeneous limit and obtain an exact
expression for the two time characteristic function of the CTRW
process in Laplace space. This main result is a generalization of the
Montroll--Weiss equation \cite{MW} which gives the characteristic
function of the single time PDF. We use our result to investigate the
validity of a multi-point fractional kinetic equation of Baule and
Friedrich \cite{BauleEPL}. We moreover obtain the two-point
correlation function $\left\langle x_1(t_1) x_2(t_2) \right\rangle$ in
a biased CTRW. We note that even this simple correlation function
cannot be found from the Green's function $p(x_1,t_1)$ of the CTRW,
since the process is non Markovian.

\section{Definitions and Notation}
We consider a standard CTRW model in one dimension with a walker
starting at the origin at time $t=0$
\cite{MW,Kehr,BouchaudREV,Metzler,Zaslavsky,Flom}. Waiting times
between jumps are independent identically distributed (IID) random
variables with a common PDF $\psi(t)$. After waiting the particle
makes a jump and the process is then renewed.  The jump lengths are
also IID random variables, with a PDF $f(\delta x)$.  The waiting
times and the lengths of jumps are mutually independent.

Let $p(x_1,t_1;x_2,t_2)=p(x_1,t_1;x_2,t_2|0,0)$ be the joint PDF of
finding a walker at $x_1$ at time $t_1>0$ and at $x_2$ at time
$t_2>0$.  We concentrate first on the corresponding multi-point
characteristic function, i.e. on the double Fourier double Laplace
transform of $p(x_1,t_1;x_2,t_2)$:
\begin{widetext}
\begin{equation} p(k_1,\lambda_1;k_2,\lambda_2) = \int_{-\infty}
^\infty {\rm d} x_1 \int_{-\infty} ^\infty {\rm d} x_2 \int_0
^{\infty} {\rm d} t_1 \int_0 ^{\infty} {\rm d} t_2 \; e^{i k_1 x_1 + i
k_2 x_2 - \lambda_1 t_1 - \lambda_2 t_2} p(x_1,t_1; x_2,t_2).
\label{eq01}
\end{equation}
\end{widetext} 
All over the article, the original functions and their transforms can
be distinguished on the ground of their variables $(x_1,t_1;x_2,t_2)$
and $(k_1,\lambda_1;k_2,\lambda_2)$ for originals and transforms,
respectively. The same holds for other functions encountered in the
text.

Let us now separate in Eq.(\ref{eq01}) the contributions corresponding
to the cases $t_1<t_2$ and $t_1>t_2$ and rewrite the integral as a sum
of the two terms:
\begin{equation} 
p(k_1,\lambda_1;k_2,\lambda_2) =
p_{<}(k_1,\lambda_1;k_2,\lambda_2) + p_{>}(k_1,\lambda_1;k_2,\lambda_2)
\end{equation} 
with
\begin{widetext}
\begin{eqnarray} p_{<}(k_1,\lambda_1;k_2,\lambda_2) &=& \int_{-\infty}
^\infty {\rm d} x_1 \int_{-\infty} ^\infty {\rm d} x_2 e^{i k_1 x_1 +
i k_2 x_2} \int_0 ^\infty {\rm d} t_1 \int_{t_1} ^\infty {\rm d} t_2
e^{ -\lambda_1 t_1 - \lambda_2 t_2} p(x_1,t_1; x_2,t_2)\label{eq02a}
\\ p_{>}(k_1,\lambda_1;k_2,\lambda_2) &=& \int_{-\infty} ^\infty {\rm d}
x_1 \int_{-\infty} ^\infty {\rm d} x_2 e^{i k_1 x_1 + i k_2 x_2}
\int_0 ^\infty {\rm d} t_1 \int_{0} ^{t_1} {\rm d} t_2 e^{ -\lambda_1
t_1 - \lambda_2 t_2} p(x_1,t_1; x_2,t_2).
\label{eq02}
\end{eqnarray}
For $t_1<t_2$ we define the elapsed time $\tau=t_2-t_1>0$ and the
corresponding displacement $\Delta=x_2-x_1$. Similarly, for $t_1>t_2$
we take $\tau=t_1-t_2$ and $\Delta=x_1-x_2$.  For $t_2>t_1$ we define the joint
PDF $g(x_1,t_1;\Delta,\tau)$ that the particle is at $x_1$ at time
$t_1$ and then experiences a displacement of size $\Delta$ during the
time interval $\tau$. Similar joint PDF for $t_2<t_1$ is denoted
with $g(x_2,t_2;\Delta,\tau)$.

To obtain $p(k_1,\lambda_1;k_2,\lambda_2)$ we use
$g(x_1,t_1;\Delta,\tau)$ and $g(x_2,t_2;\Delta,\tau)$ in the
corresponding terms of Eq. (\ref{eq02a}) and Eq. (\ref{eq02}).
Changing in Eq. (\ref{eq02a}) the variables according to
$t_2=t_1+\tau$ and $x_2 =x_1+\Delta$ one gets
\begin{eqnarray} 
p_{<}(k_1,\lambda_1;k_2,\lambda_2) &=& \int_{-\infty}
^\infty {\rm d} x_1 \int_{-\infty} ^\infty {\rm d} \Delta \int_0
^\infty {\rm d} t_1 \int_{0} ^\infty {\rm d} \tau e^{ -\lambda_1 t_1 -
\lambda_2 (t_1 + \tau) + i k_1 x_1 + i k_2 (x_1 + \Delta) }
g(x_1,t_1;\Delta,\tau) \nonumber \\ &=& g(k_1 + k_2,\lambda_1 +
\lambda_2; k_2,\lambda_2).
\end{eqnarray}
\end{widetext} 
A similar change of variables is made in
Eq.(\ref{eq02}) giving us
\begin{eqnarray} 
p(k_1,\lambda_1;k_2,\lambda_2) &=& g(k_1 +
k_2,\lambda_1 + \lambda_2;k_1,\lambda_1) \nonumber \\ &+& g(k_1 +
k_2,\lambda_1 + \lambda_2;k_2,\lambda_2).
\label{eq03}
\end{eqnarray}
Thus, our next task is to find the characteristic function $g$.  We
consider $t_1<t_2$ and soon concentrate on $g(k_1,\lambda_1;k,s)$ which is
the double Laplace and double Fourier transform of
$g(x_1,t_1;\Delta,\tau)$ according to the rule $x_1 \to k_1$, $t_1 \to
\lambda_1$, $\Delta \to k$ and $\tau \to s$.  Let $n_1$ be the random
number of jumps made by a walker during the time interval $(0,t_1)$,
$n_2$ the number of jumps made in the time interval $(t_1,t_1 +
\tau)$, and $P_{n_1,n_2}(t_1,\tau)$ the joint probability of these
random variables.  Since waiting times and jump lengths are
independent, we may write
\begin{equation} g(x_1,t_1;\Delta,\tau) = \sum_{n_1,n_2=0}^\infty
P_{n_1,n_2}(t_1,\tau) p(x_1;n_1) p(\Delta | x_1;n_2)
\label{important}
\end{equation} 
where $p(x_1;n_1)$ is the probability density to find a particle at
$x_1$ after $n_1$ steps and $p(\Delta | x_1;n_2)$ is the probability
density of the displacement $\Delta$ of a particle starting at $x_1$
after $n_2$ steps. Eq. (\ref{important}) is the key to all our further
considerations, and it shows that the problem can be divided into
three: calculation of $P_{n_1,n_2}(t_1,\tau)$ which is not trivial,
and the calculation of $p(x_1;n_1)$ and $p(\Delta | x_1;n_2)$.  The
latter two describe simple random walks in discrete time,
moreover, since the jump lengths are independent
$p(\Delta |x_1;n_2)$ does not depend directly on $x_1$.  In this case
the corresponding displacements
\begin{equation} 
x_1 = \sum_{i=1} ^{n_1} \delta x_i , \ \ \ \ \
\Delta=\sum_{i=n_1+1} ^{n_1+n_2} \delta x_{i},
\label{eq04}
\end{equation}
are sums of IID random variables, and their characteristic functions
are:
\begin{equation}
\langle e^{ i k_1 x_1} \rangle = \int_{-\infty} ^\infty p(x_1;n_1) 
e^{i k_1 x_1} {\rm d} x_1  = \langle e^{ i k_1 \delta x} \rangle^{n_1}= f^{n_1}(k_1)
\label{eqelne1}
\end{equation}
and 
\begin{equation}
\langle  e^{i k \Delta} \rangle= \int_{-\infty} ^\infty p(\Delta|x_1;n_2) 
e^{i k \Delta }{\rm d} \Delta = f^{n_2}(k) 
\end{equation}
where $f(k)=\langle \exp( i k \delta x)\rangle$ is the one step
characteristic function i.e. the Fourier transform of $f(\delta x)$.
Passing to the Fourier -- Laplace representation we hence get
\begin{equation} 
g(k_1,\lambda_1; k,s) =\sum_{n_1,n_2=0} ^\infty
P_{n_1,n_2}(\lambda_1,s) f^{n_1} (k_1) f^{n_2} (k).
\label{eq04a}
\end{equation}
where $P_{n_1,n_2}(\lambda_1,s)$ is the double Laplace transform of
$P_{n_1,n_2}(t_1, \tau)$ in its temporal variables.  We thus see that
this Laplace-transform $P_{n_1,n_2}(\lambda_1,s)$ of the probability
of the number of steps during the corresponding time intervals plays
the key role in our theory and we turn now to determining it.

\section{Statistics of Numbers of Steps $P_{n_1,n_2}(\lambda_1,s)$}
Let us now consider the set of jump times as a point process on the
time axis and let $\overline{t}_1,\overline{t}_2,\cdots \overline{t}_i
\cdots$ denote the corresponding points: $\overline{t}_1$ is the
instant of time when the first jump was made, $\overline{t}_2$ is the
time when the second jump was made etc.  As mentioned, according to
the CTRW model the waiting times $\overline{t}_1$, $\overline{t}_2-
\overline{t}_1$, $\overline{t}_3-\overline{t_2}$ etc are IID random
variables with the common PDF $\psi(t)$. The realizations of the
process with $n_1$ jumps up to time $t_1$ and $n_2$ jumps between
$t_1$ and $t_1+\tau$ are those that satisfy $\overline{t}_{n_1} < t_1
< \overline{t}_{n_1+ 1}$ and $\overline{t}_{n_1 + n_2} <t_1 + \tau <
\overline{t}_{n_1 + n_2 +1}$ respectively. We introduce the indicator
function $I(x)$ of a logical variable $x$ so that $I(x)=1$ if the
condition $x$ in the brackets holds ($x$ is true) and $I(x)=0$
otherwise. As usual the probability for $x$ to be true is then the
mean $\left \langle I(x) \right\rangle$ of $ I(x)$ over realizations.
For the case under consideration
\begin{eqnarray} 
&&P_{n_1,n_2}(t_1,\tau) = \\ &&\left \langle I \left(
\overline{t}_{n_1} < t_1 < \overline{t}_{n_1 + 1}\right) I \left(
\overline{t}_{n_1 + n_2} < t_1 + \tau < \overline{t}_{n_1 + n_2 +
1}\right)\right \rangle.  \nonumber
\label{Eq11}
\end{eqnarray}
The argument of the second indicator function can be rewritten as $(
\overline{t}_{n_1 + n_2}-t_1 < \tau < \overline{t}_{n_1 + n_2 +1}-t_1)$.  
The double Laplace transform of $P_{n_1,n_2}(t_1,\tau)$ is 
\begin{eqnarray} 
&& P_{n_1 , n_2 }(\lambda_1,s)= \int_0 ^\infty {\rm
d}\tau e^{ - s \tau} \int_0 ^\infty {\rm d} t_1 e^{ - \lambda_1 t_1}
P_{n_1,n_2}(t_1,\tau) \nonumber \\ && = \left \langle \int_0 ^\infty
{\rm d} t_1 e^{ - \lambda_1 t_1} I \left(\overline{ t}_{n_1} < t_1
<\overline{t}_{n_1 + 1}\right) \times \right. \nonumber \\ &&
\left. \int_0 ^\infty {\rm d} \tau e^{ - s \tau} I \left(
\overline{t}_{n_1 + n_2}-t_1 < \tau < \overline{t}_{n_1 + n_2 +
1}-t_1\right) \right\rangle.
\label{Eq12}
\end{eqnarray}
where we used the linearity of the Laplace transform to interchange
the sequence of integration and ensemble averaging. Note that only
the second indicator function contains $\tau$ as a variable and the
second integral in Eq.(\ref{Eq12}) 
\[
\mbox{S.I.}\equiv \int_0 ^\infty {\rm d} \tau e^{- s
\tau} I \left( \overline{t}_{n_1+n_2} - t_1< \tau < \overline{t}_{n_1
+ n_2 + 1} - t_1 \right)
\]
is rather trivial and it exhibits three behaviors:
$(i)$ $\mbox{S.I.}=0$ for $\overline{t}_{n_1 + n_2 + 1} - t_1 < 0$, $(ii)$
\[
\mbox{S.I.}=\int_0 ^{\overline{t}_{n_1 + n_2+1} - t_1} e^{ -
s \tau} {\rm d} \tau = \frac{1 - 
e^{ - s \left( \overline{t}_{n_1+n_2 + 1} - t_1\right)}}{s}
\] 
for $\overline{t}_{n_1 + n_2} <t_1 <\overline{t}_{n_1 + n_2 + 1}$ and $(iii)$
\begin{eqnarray*}
\mbox{S.I.} &=& 
\int_{\overline{t}_{n_1+ n_2} -
t_1} ^{\overline{t}_{n_1 + n_2+1} - t_1} e^{ - s \tau} {\rm d} \tau \\
&=&
{e^{ - s \left( \overline{t}_{n_1 + n_2} - t_1 \right) } - e^{ -
s \left( \overline{t}_{n_1+ n_2 + 1 } - t_1\right) } \over  s}
\end{eqnarray*}
for $t_1 < \overline{t}_{n_1 + n_2} $. 
We note that condition $(ii)$
and the condition $\overline{ t}_{n_1} < t_1 <\overline{t}_{n_1 + 1}$
can hold simultaneously only when $n_2$=0, while condition $(iii)$
and the condition $\overline{ t}_{n_1} < t_1 <\overline{t}_{n_1 + 1}$
can only hold simultaneously if $n_2 \neq 0$. Using these behaviors we
now get from Eq. (\ref{Eq12})
\begin{widetext}
\begin{eqnarray} 
P_{n_1 ,0}(\lambda_1,s) &=& \left \langle
\int_{\overline{t}_{n_1}} ^{\overline{t}_{n_1+1} } {\rm d} t_1 e^{ -
t_1 \lambda_1 } I \left( \overline{t}_{n_1} < t_1 < \overline{t}_{n_1
+ 1} \right) {1 - e^{ - s \left( \overline{t}_{n_1+ 1 } - t_1\right) }
\over s} \right \rangle \nonumber \\ & = & \left \langle { e^{
-\overline{t}_{n_1} \lambda_1} - e^{ - \lambda_1 \overline{t}_{n_1 +
1}} \over s \lambda_1 } - { e^{ - s \overline{t}_{n_1 + 1} }\over s} {
e^{ - ( \lambda_1 - s ) \overline{t}_{n_1}} - e^{ - ( \lambda_1 - s)
\overline{t}_{n_1 + 1}} \over \lambda_1 -s} \right \rangle
\label{Eq14}
\end{eqnarray} 
for $n_2=0$ and
\begin{eqnarray} 
P_{n_1 ,n_2}(\lambda_1,s) & = & \left \langle \int_0
^{\overline{t}_{n_1 + n_2} } {\rm d} t_1 e^{ - t_1 \lambda_1 } I
\left( \overline{t}_{n_1} < t_1 < \overline{t}_{n_1 + 1} \right) {e^{
- s \left( \overline{t}_{n_1 + n_2} - t_1 \right) } - e^{ - s \left(
\overline{t}_{n_1+ n_2 + 1 } - t_1\right) } \over s} \right \rangle
\nonumber \\ & = &\left \langle {{ e^{ - s \overline{t}_{n_1 +n_2 }} -
e^{ - s \overline{t}_{n_1 +n_2+ 1}}} \over s} { e^{ - ( \lambda_1 - s
) \overline{t}_{n_1}} - e^{ - ( \lambda_1 - s) \overline{t}_{n_1 + 1}}
\over \lambda_1 -s} \right \rangle \label{Eq15}
\end{eqnarray} 
for $n_2 \neq 0$.
\end{widetext}

Since waiting times are IID random variables, one has
\begin{equation} 
\left \langle e^{ - \overline{t}_{n_1} \lambda_1 }
\right \rangle = \psi^{n_1} \left( \lambda_1 \right)
\end{equation} 
and
\begin{eqnarray} 
\left \langle e^{ -s \overline{t}_{n_1 + 1}} e^{ - (\lambda_1 - s)
\overline{t}_{n_1}} \right \rangle &=& \left \langle e^{ - \lambda_1
\overline{t}_{n_1}} e^{ -s (\overline{t}_{n_1 +
1}-\overline{t}_{n_1})} \right \rangle \nonumber \\ 
&=& \psi^{n_1} (\lambda_1) \psi(s). 
\label{Eq16}
\end{eqnarray}
Similar expressions hold also 
for other terms in Eqs. (\ref{Eq14}) and (\ref{Eq15}). 
Here $\psi\left( \lambda_1 \right)$ and $\psi(s)$ are Laplace transforms
of the waiting time PDF $\psi(t_1)$ and $\psi(\tau)$ respectively.
Using these expressions we get:
\begin{equation} P_{n_1 ,0}\left(\lambda_1,s \right) = 
{\psi^{n_1} \left( \lambda_1 \right) \over s} \left[ { 1 -
\psi(\lambda_1) \over \lambda_1} - {\psi(s) -
\psi(\lambda_1) \over \lambda_1-s} \right]
\label{Eq17}
\end{equation}
and
\begin{eqnarray} 
&&P_{n_1 , n_2 } \left( \lambda_1,s \right) =  \label{Eq18} \\
&& {\psi^{n_1} \left( \lambda_1 \right) \psi^{n_2 - 1} \left(
s \right) \over s } \left[ 1 -\psi\left( s \right) \right] 
{{\left[\psi \left( s \right) - \psi \left( \lambda_1 \right) \right]} 
\over {\left( \lambda_1 - s \right)}} 
\nonumber 
\end{eqnarray}
for $n_2 \ge 1$. Note that Eqs. (\ref{Eq17},\ref{Eq18}) give the proper normalization
since $\sum_{n_1=0}^\infty \sum_{n_2=0} ^\infty P_{\lambda_1,s} ( n_1,n_2) = 1/ ( \lambda_1 s)$.

We now consider limiting behaviors of Eqs. (\ref{Eq17},\ref{Eq18}).
The probability of making no steps in the time interval
$(t_1,t_1+\tau)$ is given in double Laplace representation by
\begin{equation} 
\sum_{n_1=0} ^\infty P_{n_1 , 0}(\lambda_1,s) =
{ 1 \over s \lambda_1 } - {\psi(s) -\psi (\lambda_1) \over
s \left( \lambda_1 - s \right) \left[ 1 - \psi\left( \lambda_1
\right) \right]},
\label{Eq19}
\end{equation}
which was obtained previously \cite{Godreche}. Let $t_f$ be the time
between $t_1$ and the first jump event after $t_1$:
$t_f=\overline{t}_{n_1 + 1} -t_1$.  The random variable $t_f$ is
sometimes called the forward recurrence time. Let its PDF be given by
$h(t_f;t_1)$ depending on $t_1$ as a parameter.  In double Laplace
representation $t_1 \to \lambda_1$ and $t_f \to s$ one finds
\cite{Godreche,Dynkin}
\begin{equation} h\left( s; \lambda_1 \right) = {\psi(s) -
\psi(\lambda_1) \over ( \lambda_1 - s) \left[ 1 -
\psi(\lambda_1) \right]},
\label{Eq20}
\end{equation}
as follows from Eq. (\ref{Eq19}) by noting that the probability of making no
jump in the time interval $(t_1,t_1 + \tau)$ is $1 - \int_0 ^{\tau} h(t_f;t_1) {\rm d} t_f$. 
The probability of making $n_2 \ge 1$ jumps in $(t_1,t_1 + \tau)$ is
\begin{equation} 
\sum_{n_1=0} ^\infty P_{n_1,n_2}(\lambda_1,s) =
h(s;\lambda_1) { 1 - \psi(s) \over s} \psi^{n_2 - 1}(s).
\label{Eq21}
\end{equation}
In the space of originals this equation corresponds to a convolution
of the PDF of forward recurrence time $t_f$ with the PDFs of the the
following $n_2 -1$ waiting times; the factor $[1-\psi(s)]/s$ comes
from the probability of not jumping between the last event in the
sequence and the end of observation at $t_2$.

\section{Two-point characteristic functions}
We are now able to find the characteristic function 
$g(k_1,\lambda_1;k,s)$, Eq. (\ref{eq04a}), using Eqs. (\ref{Eq17},\ref{Eq18}):
\begin{widetext}
\begin{eqnarray}
g(k_1,\lambda_1;k,s) &=& 
\left[ {{1 - \psi \left( \lambda_1 \right) }  \over {\lambda_1 s} } 
-{{\psi \left(s\right) - \psi \left(\lambda_1 \right)} 
\over s \left( \lambda_1 - s \right) } \right]
{ 1 \over 1 - \psi\left( \lambda_1 \right) f \left( k_1 \right)}  \nonumber \\
&+& {f\left( k \right) \left[ 1 - \psi\left( s \right) \right]
\left[ \psi \left( s \right) - \psi \left( \lambda_1
\right) \right] \over s \left( \lambda_1 - s \right) }
{ 1 \over 1 -
\psi \left( \lambda_1 \right) f \left( k_1 \right) } { 1 \over 1
- \psi (s) f\left( k \right) } \nonumber \\
&=& \left[\frac{1}{\lambda_1 s}-\frac{h(s;\lambda_1)}{s}\right]
\frac{1-\psi(\lambda_1)}{1-\psi(\lambda_1)f(k_1)} +
\frac{1-\psi(\lambda_1)}{1-\psi(\lambda_1)f(k_1)} \frac{h(s;\lambda_1)f(k)}{s} 
\frac{1-\psi(s)}{1-\psi(s)f(k)}. 
\label{Eq25}
\end{eqnarray}
\end{widetext}
This equation can be written in a more transparent way. First we
recall the Montroll--Weiss equation.  For a CTRW starting at time
$t=0$, the single point PDF $P_{{\rm MW}}(x,t)$ of
the particle being at site $x$ at time $t$, is given in Laplace $t
\to \lambda$ Fourier $x \to k$ space in terms of the
Montroll--Weiss equation \cite{MW}
\begin{equation} 
P_{{\rm MW}}\left[\psi(\lambda)
,f(k)\right] = {1 - \psi (\lambda) \over \lambda} 
{1 \over 1 - f(k) \psi(\lambda)}.
\label{Eq26}
\end{equation}
The Montroll--Weiss equation explicitly assumes that the waiting
time for the first step has the same PDF as all further waiting
times. On the other hand, one may consider
situations where the waiting time PDF for the first step, $\psi_1(t)$
differs from the PDFs of all other waiting times $\psi(t)$
\cite{Tunaley}. The single point PDF describing this more general
process \cite{ACTRW,ACTRWPRL} called aging random walk \cite{remark} 
is denoted by $P_{{\rm ARW}}(x,t)$. In the Laplace-- Fourier representation $x \to
k$, $t \to \lambda$ one finds
\begin{eqnarray} 
&& P_{{\rm ARW}}\left[\psi_1(\lambda), \psi(\lambda)
,f(k)\right] = \label{Eq27} \\
&& \qquad {1 - \psi_1(\lambda) \over \lambda}
+ {\psi_1 (\lambda) f(k) \over 1 - f(k)
\psi(\lambda_1)} { 1 - \psi(\lambda) \over \lambda}. \nonumber
\end{eqnarray}
The aging random walk reduces to the Montroll--Weiss
CTRW if $\psi_1(t) = \psi(t)$. Using Eqs. (\ref{Eq26}) and Eq.(\ref{Eq27}) one can
rewrite Eq. (\ref{Eq25}) as
\begin{eqnarray} 
&& g(k_1,\lambda_1;k,s) = \label{Eq28} \\
&& P_{{\rm MW}}\left[ \psi
\left( \lambda_1 \right), f \left( k_1 \right) \right] P_{{\rm ARW}}
\left[ \lambda_1 h(s;\lambda_1) , \psi \left( s \right) ,
f\left( k \right) \right]. \nonumber
\end{eqnarray}
We see that the solution for $g$ corresponds to a convolution of two
PDFs, the one of the Montroll--Weiss CTRW and the one of the aging
CTRW with the first waiting time PDF formally put to $\psi_1(\tau) =
dh(\tau,t)/d t$. We note that the fact that the final characteristic
function is a convolution and not a simple product of
$P_{{\rm MW}}(x_1,t_1)$ and $P_{{\rm ARW}}(\Delta,\tau; t_1)$,
as found for simple
Markovian diffusion, has to do with the correlations between $x_1$ and
$x_2$ which in turn is related to the correlation between $n_1$ and
$n_2$.  These arise through subtle correlations between the number of
steps $n_1$ and the forward recurrence time.

After getting $g(k_1,\lambda_1;k,s)$ we can turn to the original
characteristic function. Using Eqs. (\ref{eq03}, \ref{Eq28}) we find

\begin{widetext}
\begin{equation} 
p(k_1,\lambda_1;k_2,\lambda_2) = \sum_{i=1,2} P_{{\rm
MW}} \left[ \psi(\lambda_1 + \lambda_2),f(k_1 +
k_2)\right]P_{{\rm ARW}} \left[ \left(\lambda_1 + \lambda_2\right)
h(\lambda_1 + \lambda_2; \lambda_i ) , \psi\left(
\lambda_i \right), f\left( k_i \right) \right].
\label{eqMain}
\end{equation}
\end{widetext}
One can check that if $k_1=0$ or $k_2=0$ (i.e. integrating the overall
distribution over $x_1$ or $x_2$ respectively) we recover
the Montroll--Weiss equation (\ref{Eq26}) for a one-point
characteristic function, for example
\begin{equation} 
p\left(k_1=0,\lambda_1;k_2,\lambda_2\right) = {1
\over \lambda_1 \lambda_2} {1 - \psi \left( \lambda_2 \right)
\over 1 - \psi\left( \lambda_2 \right) f(k_2) }
\end{equation}
as it should.

We now investigate the continuum limit of our main equation (\ref{eqMain}),
corresponding to long $t_1$ and $t_2$, using the standard
long wave length (small $k$) and small frequency approximation
\cite{Metzler}. We consider first the non-biased random walks with a
finite second moment of jump lengths $\langle \delta x^2 \rangle$,
which means that for small $k$
\begin{equation} 
f\left( k \right) \sim 1 - {\langle \delta
x^2 \rangle k^2 \over 2}+\cdots .
\end{equation}
For $\lambda_1 \to 0$
\begin{equation} 
\psi(\lambda_1) \sim 1 - A \left(
\lambda_1\right)^\alpha
\label{eqPSI}
\end{equation}
where $0<\alpha\le 1$ and $A>0$. For the special case $\alpha=1$, $A$
is the mean waiting time. This case corresponds to asymptotically normal
diffusion. If $\alpha<1$ the mean time
between jumps diverges, which leads to anomalous behaviors.
In this limit
\begin{widetext}
\begin{eqnarray}
p(k_1,\lambda_1;k_2,\lambda_2) && \sim
{ \left( \lambda_1 + \lambda_2\right)^{\alpha -1}
\over \left( \lambda_1 + \lambda_2\right)^\alpha + D_{\alpha} |k_1 +
k_2|^2 } 
\left\{ \sum_{i=1,2} { \left( \lambda_1 + \lambda_2\right)
(\lambda_i)^\alpha - \lambda_i\left( \lambda_1 +
\lambda_2\right)^\alpha \over \lambda_1 \lambda_2 \left( \lambda_1 +
\lambda_2\right)} + \right. 
\nonumber \\
&&    + \left( \lambda_1 + \lambda_2 \right)
{{\left(\lambda_1 + \lambda_2 \right)^\alpha - (\lambda_i)^\alpha} \over
{\lambda_{\overline{i}}\left( \lambda_1 + \lambda_2 \right)^\alpha}} 
{(\lambda_1)^{\alpha-1} \over {(\lambda_i)^\alpha + D_{\alpha} (k_i)^2}} 
\Bigg\}
\label{eqMain1}
\end{eqnarray}
where $\overline{1} = 2$ and $\overline{2} = 1$ and $D_\alpha=\langle
\delta x^2 \rangle / ( 2 A)$ is the fractional diffusion constant
\cite{BarkaiPRE}.  Previously Baule and Friedrich \cite{BauleEPL}
wrote a multi-point fractional diffusion equation for
$p(x_1,t_1;x_2,t_2)$, whose solution in Laplace-Fourier space is
exactly Eq.(\ref{eqMain1}).
\end{widetext}

\section{Correlation function for Biased CTRW}
We now consider the simplest correlation function
\begin{equation} \langle x_1(\lambda_1) x_2(\lambda_2) \rangle = -
{\partial \over \partial k_1} {\partial \over \partial k_2 }
p(k_1,\lambda_1;k_2,\lambda_2)|_{k_1=k_2=0}.
\label{eqCo1}
\end{equation}
For a biased CTRW with finite variance of jump lengths, the small $k$
expansion reads
\begin{equation} 
f(k) \sim 1 + i \langle \delta x \rangle k -
{\langle \delta x^2 \rangle k^2 \over 2} + \cdots,
\label{eqCo2}
\end{equation}
where $\langle \delta x \rangle$ is the mean step length.  Using
Eqs. (\ref{eqMain},\ref{eqCo1},\ref{eqCo2})
\begin{eqnarray} 
\langle x_1(\lambda_1) x_2 (\lambda_2) \rangle &=&
{\langle \delta x^2 \rangle \over \lambda_1 \lambda_2 \left[ 1 -
\psi\left( \lambda_1 + \lambda_2\right)\right]} \nonumber \\
&+& { 2 \langle
\delta x\rangle^2 \over \lambda_1 \lambda_2 \left[ 1 - \psi
\left( \lambda_1 + \lambda_2 \right) \right]^2 } \label{eqcor3} \\
&+& {\langle \delta
x\rangle^2 \over 1 - \psi \left( \lambda_1 + \lambda_2\right) }
\sum_{i=1,2} { h(\lambda_1+\lambda_2; \lambda_i ) \over
\lambda_i \left[ 1 - \psi \left( \lambda_i \right)\right]}. \nonumber
\end{eqnarray}
Now we pass to the small $\lambda_1$ and $\lambda_2$ limit and  consider
the scaling limit of large $t_1$ and $t_2$ when their ratio is
arbitrary, using Eqs. (\ref{eqPSI},\ref{eqcor3})
\begin{eqnarray}
\langle x_1(\lambda_1) x_2 (\lambda_2) \rangle & \sim&
{\langle \delta x^2 \rangle \over A \lambda_1 \lambda_2 \left(
\lambda_1 + \lambda_2 \right)^\alpha} + \nonumber \\
& +& {\langle \delta x \rangle^2
\over A^2 } { 1/(\lambda_1)^\alpha + 1 / (\lambda_2)^\alpha \over
\lambda_1 \lambda_2 \left( \lambda_1 + \lambda_2 \right)^\alpha}.
\label{eqcor4}
\end{eqnarray}
Laplace transform of Eq. (\ref{eqcor4}) is found using the approach 
discussed in Appendix B of Ref.\cite{BaulePRE}. For $t_2>t_1$ one has:
\begin{eqnarray} 
&&\langle x_1 (t_1) x_2 (t_2) \rangle \sim {\langle
\delta x^2 \rangle \over A} { (t_1)^\alpha \over \Gamma(1 + \alpha) } + 
\label{eqcor5} \\
&& {\langle \delta x \rangle^2 \over A^2} \left[ { (t_1)^{2 \alpha}
\over \Gamma\left( 1 + 2 \alpha \right) }
+ {(t_1 t_2)^\alpha \over \Gamma\left( 1 +
\alpha \right)^2 }  F\left(
\alpha,-\alpha,\alpha+1;{t_1 \over t_2}\right) \right]. \nonumber
\end{eqnarray}
Taking $t_1>t_2$ corresponds to simple interchange of the arguments. 
Here $F(a,b;c;z)$ is a hypergeometric function \cite{Abramowitz}.
Let us check limiting behaviors of the correlation function 
Eq. (\ref{eqcor5}).
For an unbiased process $\langle \delta x \rangle=0$ 
the first term on the right hand side of equation
(\ref{eqcor5}) is the only non-vanishing term
\begin{equation}
\langle x_1 (t_1) x_2 (t_2) \rangle \sim {\langle
\delta x^2 \rangle \over A} { (t_1)^\alpha \over \Gamma(1 + \alpha) }
\label{eqtriv}
\end{equation}
which represents the dispersion of the walker's positions at time $t_1$. 
On the other hand, for the biased CTRW $\langle \delta x \rangle \ne 0$
the leading behavior is given by the second line
in Eq. (\ref{eqcor5}) and we may neglect the first term.
Several limiting situations can be used as benchmarks.  
For $t_2 \gg t_1$ one has $F(\alpha,-\alpha;\alpha+1;x)\sim 1 - \alpha^2 x/(1 + \alpha) +
O(x^2)$ so that
\begin{equation} 
\langle x_1 (t_1) x_2 (t_2) \rangle \sim \langle
x_1(t_1) \rangle \langle x_2 (t_2) \rangle 
\label{decouple}
\end{equation}
with
\begin{eqnarray}
&& \langle x_1(t_1)
\rangle \sim {\langle \delta x \rangle \over A} {(t_1)^\alpha \over
\Gamma(1 + \alpha) } \\
&& \langle x_1(t_2) \rangle \sim {\langle
\delta x \rangle \over A} {(t_2)^\alpha \over \Gamma(1 + \alpha) },
\nonumber
\end{eqnarray}
which proves the decoupling of correlations for $t_2 \gg t_1$.  
In the opposite limit of $t_2 \to t_1$ one uses Eq.(15.3.6) 
of Ref.(\cite{Abramowitz}) 
and $F(a,b;c;0)=1$ 
to get 
$F(\alpha,-\alpha;\alpha+1;1)=[\Gamma(\alpha+1)]^2/\Gamma(2\alpha+1)$ and 
to obtain
\begin{equation} 
\lim_{t_2 \to t_1} \langle x_1 (t_1) x_2 (t_2)
\rangle \sim {\langle \delta x^2 \rangle \over A} { (t_1)^\alpha \over
\Gamma(1 + \alpha) } + {2 \langle \delta x \rangle^2 (t_1)^{2 \alpha} \over \Gamma(1 +
2 \alpha) A^2},
\label{eqcor6}
\end{equation}
which is the mean square displacement in the biased CTRW
\cite{Shlesinger}, as expected.

Another limit is the Markovian case $\alpha=1$, for which
$F(1,-1;2;x)=1-x/2$.  In this case the decoupling,
Eq.(\ref{decouple}), is valid at all times. The fact that for $\alpha
< 1$ Eq.(\ref{decouple}) holds only for $t_2 \gg t_1$, 
i.e. the existence of
nontrivial correlations between $x_1$ and $x_2$, has to do with the
correlations between the number of steps before and after the first
observation time $t_1$ which we discussed in Sec. IV.  Thus, if
relatively few jumps take place during the time interval $(0,t_1)$,
i.e. $n_1 \ll \langle n_1 \rangle$, the typical
displacement $x_1$ is inevitably small, and then
the particle is likely to be
effectively trapped at its position at $t_1$ for a very long time 
which is of the order of $t_1$.  In this case also the forward
recurrence time is long in statistical sense. This implies that also $n_2$ is 
going to be relatively small, since the particle will likely  wait
for a long time for its first step after $t_1$, which leads to small absolute
values of $x_2$ as well. Hence correlations for $\alpha<1$
are built even when
$t_1$ and $t_2$ are very long.

\section{Discussion}

The Montroll--Weiss equation (\ref{Eq26}) expresses the characteristic
function of the CTRW in terms of Laplace and Fourier transforms of the
PDFs of the waiting times and jump lengths. Similarly,
Eq. (\ref{eqMain}) gives the two dimensional characteristic function
of the CTRW process. From this equation we may derive two dimensional
correlation functions for the CTRW process, for example we considered
the biased CTRW. We showed that the two dimensional characteristic
function depends on the probability of $n_1$ renewals in $(0,t_1)$ and
$n_2$ renewals in $(t_1,t_2)$ (for $t_2 > t_1)$ and that these numbers
of steps are correlated. For characteristic functions of order $N$
higher than two one would have to calculate renewal statistics in $N$
intervals.  In principle this calculation can be performed using the
same technique we used here for example to calculate
$P_{n_1,n_2,n_3}(t_1,t_2,t_3)$ etc.

The two dimensional characteristic function is shown to be related to
the Montroll--Weiss and aging CTRW single point characteristic
functions. Thus even though the process is non-Markovian information
on one dimensional characteristic functions is sufficient to find the
two dimensional characteristic function. This simplification is
obviously related to the renewal property of the underlying random
walk.  Finally, starting with the CTRW model we derived the solution
of the multi-point fractional diffusion equation \cite{BauleEPL}, in
Fourier-Laplace space, thus giving further justification for this new
equation.

{\bf Acknowledgment} EB thanks the Israel Science Foundation for
support and 
R. Friedrich for discussions.


\begin{thebibliography}{99}





\bibitem{MW} E.W. Montroll, and G. Weiss, {\em J. Math. Phys.} {\bf
6}, 167 (1965).

\bibitem{Kehr} J. W.  Haus, and K. W. Kehr, {\em Phys. Rep.} {\bf 150}
263 (1987).

\bibitem{BouchaudREV} J.P. Bouchaud and A. Georges, {\em Phys. Rep.} 
{\bf 195} 127 (1990).

\bibitem{Metzler} R. Metzler, J. Klafter, {\em Phys. Rep.} {\bf 339} 1
(2000).

\bibitem{Zaslavsky} G. M. Zaslavsky, {\em Phys. Rep.} {\bf 371} 461
(2002).

\bibitem{Flom} O. Flomenbom, and J. Klafter, {\em Phys. Rev. Lett.}
{\bf 95} 098106 (2005).

\bibitem{FFPEPRL} R. Metzler, E. Barkai, and J. Klafter {\em
Phys. Rev. Lett.} {\bf 82} 3563 (1999)

\bibitem{BarkaiPRE} E. Barkai, R. Metzler, and J. Klafter {\em
Phys. Rev E} {\bf 61} 132 (2000)

\bibitem{Bel} G. Bel, E. Barkai {\em Phys. Rev. Lett.} {\bf 94} 240602
(2005).

\bibitem{ACTRW} E. Barkai, Y. C. Cheng, {\em J. of Chemical Physics}
{\bf 118} 6167 (2003).

\bibitem{ACTRWPRL} E. Barkai {\em Phys. Rev. Lett.} {\bf 90} 104101
(2003).

\bibitem{SokolovPRL} I.M. Sokolov, and J. Klafter {\em
Phys. Rev. Lett.} {\bf 97} 140602 (2006).

\bibitem{ecm} E. Barkai, cond-mat/0608155.

\bibitem{Grigolini} P. Allegrini, P. Grigolini, L. Palatella,
B. J. West {\em Phys. Rev.  E} {\bf 70} 046118 (2004).

\bibitem{Barsegov} V. Barsegov, S.  Mukamel, {\em J. Phys. Chem. A}
{\bf 108} 15 (2004).

\bibitem{Mukamel} F. Sanda, S.  Mukamel. {\em Phys Rev E} {\bf 72}
031108 (2005).

\bibitem{BaulePRE} A. Baule, and R. Friedrich {\em Physical Review E}
{\bf 71} 026101 (2005)

\bibitem{BauleEPL} A. Baule, and R. Friedrich {\em Europhysics
Letters} {\bf 77} 10002 (2007).

\bibitem{GraKla} R. Granek, J. Klafter {\em Phys. Rev. Lett.} {\bf 95} (9) 098106 (2005) 

\bibitem{Yang} H. Yang et al {\em Science} {\bf 302} (2003) 262.

\bibitem{Min} W. Min et al {\em Phys. Rev. Lett.} {\bf 94} (2005)
198302.


\bibitem{Margolin} G. Margolin, E. Barkai, {\em J. of Chem. Phys.} {\bf
121} 1566 (2004).

\bibitem{Godreche} C. Godreche, and J. M. Luck, {\em J. of Statistical
Physics} {\bf 104} 489 (2001).

\bibitem{Dynkin} For long tailed $\psi(t)$ the PDF $h_{t_1}(E)$ is
given according to a limit theorem. E. B. Dynkin {\em Selected
Translations in Mathematical Statistics and Probability} (American
Mathematical Society, Providence 1961) Vol. 1 p. 249 (see


\cite{Godreche},\cite{ACTRW} for details).


\bibitem{Tunaley} J.K.E. Tunaley {\em Phys. Rev. Lett.} {\bf 33} 1037 (1974). 

\bibitem{remark} Subdiffusive CTRW exhibits aging behaviors \cite{ACTRW}.
 Eq. (\ref{Eq27}) is used for the mathematical description
of such random walks. 
In aging CTRW  $\psi_1(t)$ depends on the age of the process. The
genuine 
aging CTRW \cite{ACTRW} 
corresponds to the choice $\psi_1(t) = h(t,t_w)$ with $t_w$ being the
time elapsing between the start of the process and the beginning of
observations (aging time). 

\bibitem{Abramowitz} M. Abramowitz and C. A. Stegun (Editors)
Handbook of Mathematical Functions (Dover, New York) 1972.

\bibitem{Shlesinger} M. F. Shlesinger, {\em J. Stat. Phys.} {\bf 10}
421 (1974).

\end{thebibliography}
\end{document}